\documentclass[11pt]{article}

\newif\ifsubmit
\submitfalse

\usepackage{scicite}
\usepackage{hyperref}
\usepackage{times}

\usepackage{amsmath}
\usepackage{amsfonts}
\usepackage{amssymb}
\usepackage{graphicx}

\usepackage{lineno}
\ifsubmit
    \linenumbers
\fi

\newenvironment{sciabstract}{%
\begin{quote} }
{\end{quote}}

\newcommand{\dtwomin}{D^2_\text{min}}

\title{Mechanical annealing and memories \\ in a disordered solid} 

\author
{Nathan C.\ Keim,${}^{1,2\ast}$ Dani Medina${}^{2}$\\
\\
\normalsize{${}^{1}$Department of Physics, Pennsylvania State University,}\\
\normalsize{University Park, PA 16802, USA}\\
\normalsize{${}^{2}$Department of Physics, California Polytechnic State University}\\
\normalsize{San Luis Obispo, CA 93407, USA}\\
\\
\normalsize{$^\ast$To whom correspondence should be addressed; E-mail:  keim@psu.edu.}
}

\date{}

\begin{document} 


\maketitle 

\begin{sciabstract}
 Shearing a disordered or amorphous solid for many cycles with a constant strain amplitude can anneal it, relaxing a sample to a steady state that encodes a memory of that amplitude. This steady state also features a remarkable stability to amplitude variations that allows one to read the memory. 
 Here we shed new light on both annealing and memory, by considering how to mechanically anneal a sample to have as little memory content as possible. 
 In experiments, we show that a ``ring-down'' protocol reaches a comparable steady state, but with no discernible memories and minimal structural anisotropy.
 We introduce a method to characterize the population of rearrangements within a sample, and show how it connects with the response to amplitude variation and the size of annealing steps. These techniques can be generalized to other forms of glassy matter and a wide array of disordered solids, especially those that yield by flowing homogeneously.
\end{sciabstract}

\section*{Introduction}

\begin{figure} 
    \centering
    \includegraphics[width=3.05in]{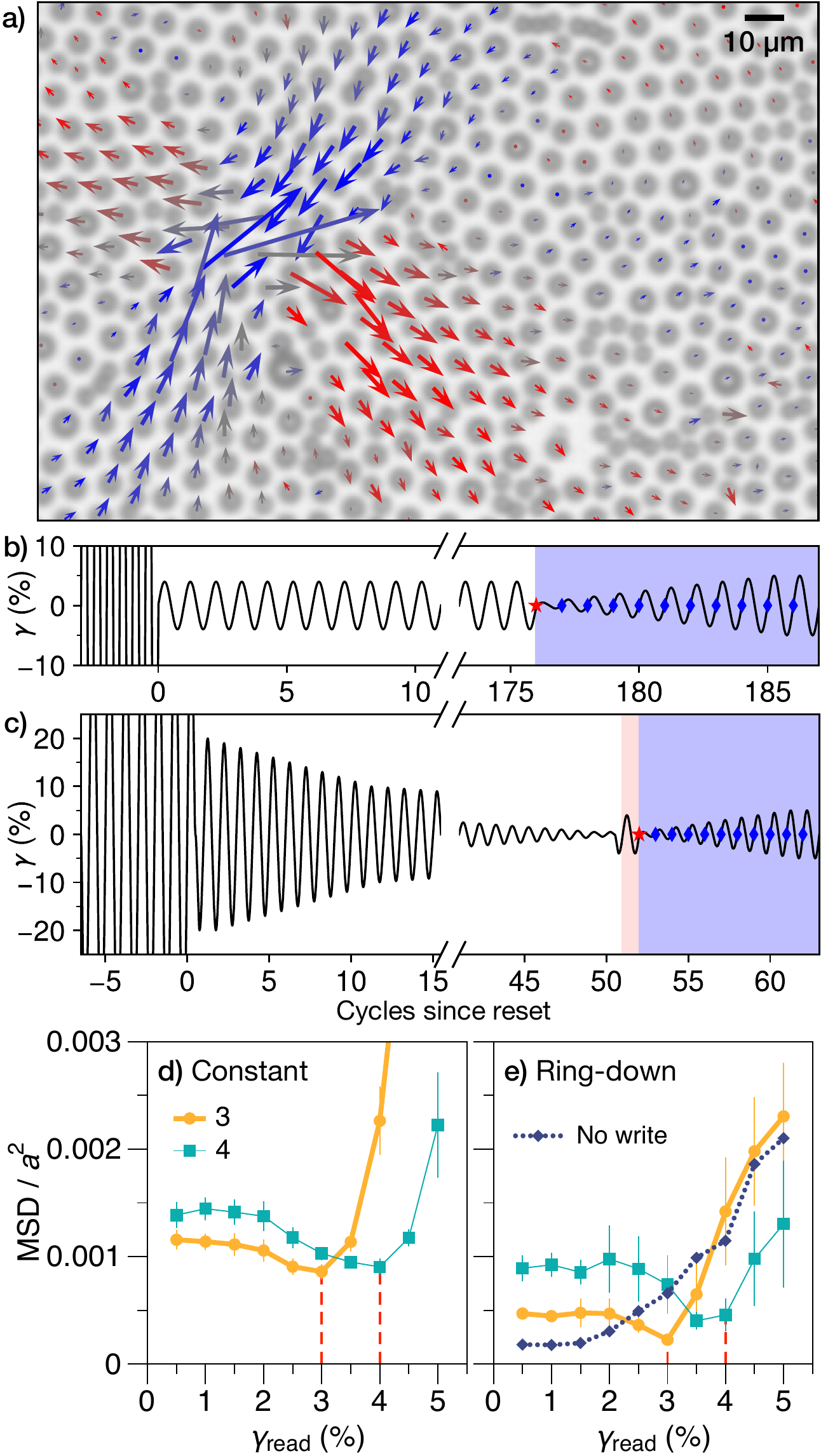}
    \caption{Annealing and single memories. 
    \textbf{(a)} 2D disordered solid, made of particles at oil-water interface with electrostatic repulsion. Arrows show displacements (exaggerated 10$\times$) at typical ``soft spot'' where particles rearrange under horizontal shear. Arrow colors represent direction.
    \textbf{(b)} Constant-amplitude shear strain ($\gamma$) protocol: after large-amplitude ``reset,'' 176 cycles anneal material and form memory of strain amplitude. Readout (blue shaded region) tests whether the annealed state (red star) can be restored, by comparing it with the states after cycles of increasing amplitude $\gamma_\text{read}$ (blue dots).
    \textbf{(c)} Ring-down annealing is followed optionally by writing of a memory (red shaded region), then readout.
    \textbf{(d, e)} Mean-squared displacement of particles, normalized by particle spacing $a$, measured during readout as $\gamma_\text{read}$ is increased. For each protocol, dips reveal written memories of 3\% or 4\% by restoring the pre-readout states. Vertical dashed lines mark expected memory values. ``No write'' curve in (e) represents no added memories and is possible only with ring-down.
    Error bars represent standard deviation of mean for multiple trials. Constant-amplitude data are from experiments in Ref.~\cite{keim2020}. 
    }
    \label{fig:simple}
\end{figure}

The steps to prepare a solid material for use typically go far beyond forming its chemical constituents into a desired shape. Techniques such as quenching (i.e., rapid cooling) or thermal annealing (slow, staged cooling) and mechanical deformation can dramatically affect the final microscopic structure, for example to increase hardness or strength~\cite{verlinden2007}. Mechanical methods show particular promise for varying the properties and broadening the applications of amorphous or disordered solids. These materials consist of atoms, particles, drops, or bubbles with negligible long-range order (Fig.~\ref{fig:simple}a), placing them far from any ground state and ensuring a strong dependence on history. For example, despite desirable properties such as corrosion resistance and large strains before failure, bulk metallic glasses tend to start life as brittle materials that fail catastrophically~\cite{schroers2013,greer2013}, but can be made less or even more brittle through deformation~\cite{sun2016}. Likewise, when studying or using the kinds of foams, concentrated suspensions, or concentrated emulsions found in consumer products or pharmaceuticals, one wishes to erase---or exploit---the effects of prior handling on rheology and microscopic structure~\cite{macosko1994,larson1998,kim2017}.

Experiments and simulations that cyclically shear disordered solids have illuminated this mechanical history-dependence, aspects of which seem to be common to 2D and 3D disordered solids with many kinds of microscopic physics~\cite{cubuk2017,regev2013a,keim2013,keim2014,royer2015,priezjev2013a,fiocco2014,sun2016,nagamanasa2014}. 
In particular, athermal simulations show that when the material starts in a higher-energy configuration---as if quenched from a high-temperature liquid, without thermal annealing---and the amplitude of shear is below a critical value, this \emph{mechanical annealing} lowers the overall energy of the structure~\cite{sun2016,regev2013a,royer2015,Das:2018vv,schinasi-lemberg2020,yeh2020,priezjev2021,bhaumik2019,sastry2021,liu2020}. Eventually, the material reaches a steady state in which the plastic rearrangements (Fig.~\ref{fig:simple}a) are fully reversible---the rearrangements become periodic, and the particle trajectories become closed loops~\cite{lundberg2008,regev2013a,keim2014,priezjev2013a,regev2015}. Since 
deformation can access a vast set of metastable arrangements of particles in these glassy materials, one might imagine that this precisely periodic behavior can be sustained only by driving at constant amplitude. Indeed, reducing the amplitude of shear strain $\gamma_0$ for even one cycle generally leaves the particles in different positions. However, these mechanically annealed solids have an additional, surprising property: resuming the previous amplitude restores the positions and the dynamics nearly perfectly~\cite{adhikari2018a,keim2020,mungan2019,regev2021}.

This kind of stability under amplitude variation lets us see that annealing has formed a memory: 
without knowing its past, one can probe a sample to reveal the strain amplitude $\gamma_a$ that was used to anneal it~\cite{fiocco2014,adhikari2018a,mukherji2019,keim2020}. We show this in Figs.~\ref{fig:simple}[b, d] with experiments on the 2D disordered solid from Fig.~\ref{fig:simple}a. We apply a series of cycles with increasing amplitude $\gamma_\text{read}$, starting with a small value and ending past $\gamma_a$ (Fig.~\ref{fig:simple}b). At the end of each readout cycle, the positions of the particles are compared with those at the end of annealing; in Fig.~\ref{fig:simple}d this is done with the mean squared displacement (MSD), normalized by the typical spacing between particles $a$ squared. At small $\gamma_\text{read}$ this displacement grows, but it drops near $\gamma_\text{read} = \gamma_a$ when the annealed state of the system is recovered. This observation and others approximate a generic behavior called return-point memory~\cite{barker1983,sethna1993,keim2019} that seems to be a property of annealed samples~\cite{keim2020,mungan2019,regev2021}. The material's tendency to retain readable memories---like many systems that do not relax to equilibrium~\cite{keim2019}---represents an opportunity for programming, adaptation, and diagnostics, but a challenge for annealing. 
If annealing is meant to relax a material and ultimately lower the entropy of its structure, this outcome would seem to involve as few discernible memories of the past as possible. Could we better understand mechanical annealing if we set out to prepare a blank slate?

In the experiments described in this paper, we prepare a disordered solid with a ``ring-down'' protocol that gradually decreases the strain amplitude $\gamma_0$ (Fig.~\ref{fig:simple}c), reminiscent of slowly cooling a material from a liquid state. 
As desired, our method leaves no memory of an annealing amplitude, and yet we can show that it still creates the conditions for return-point memory. 
The memory of an amplitude that was formed by prolonged constant-amplitude annealing turns out to be distinct from return-point memories that can be written with single cycles, and here we disentangle them.
We show that another desirable outcome of annealing is to make the material more isotropic, and that ring-down annealing does this well, erasing a memory of a shear direction from \emph{before} annealing. Finally, we introduce a more rigorous multi-cycle test of the response to amplitude variation, and show how a model links those results to details of rearrangements in a single cycle of shear. When applied to a finite-size material, this model paradoxically suggests that annealing with oscillatory shear achieves the least memory content by forming every possible memory. Our results point to generic mechanical ways to prepare, program, and probe athermal glassy matter.

\section*{Results}

The material shown in Fig.~\ref{fig:simple}a is a monolayer of polystyrene particles adsorbed at an oil-water interface, with two diameters to inhibit crystalline order. Because of the particles' long-range electrostatic repulsion~\cite{masschaele2010}, each particle is mechanically over-constrained by its neighbors but does not touch them, so that particles form a soft, frictionless jammed 2D solid (see details of samples and apparatus in Materials and Methods).
We shear the material at 0.05~Hz in a custom interfacial rheometer~\cite{shahin1986,brooks1999,reynaert2008,keim2013,keim2014,tajuelo2016} that lets us synchronously track $\sim$24,000 particles. Deformations are approximately quasistatic relative to the timescale of rearrangements.
Our analysis is based on shear strain obtained directly from particle positions, $\gamma_\text{aff}$, but for clarity we report the nominal values (e.g.~$\gamma = 3\%$, $-3.5\%$) unless otherwise noted. We find that $\gamma / \gamma_\text{aff} \sim 0.88$ 
due to the elastic modulus of the material.

Each experiment begins by shearing with a large strain amplitude $\sim$50\% for several cycles to ``reset'' the material and erase memories and prior annealing (e.g.\ Fig.~\ref{fig:simple}[b, c]). Visual inspections of the material 
confirm that this effectively replaces the visible sample with a new one, which we then anneal.
The ring-down annealing protocol (e.g.\ Fig.~\ref{fig:simple}c) is meant to gradually ``cool'' the material from plastic flow at the largest strain amplitude ($\gamma_0 = 20\%$) to a quiescent elastic solid at the smallest amplitude ($\gamma_0 = 0.25\%$). Prior experiments showed that for this system, repeating steady states and memory formation happen at $\gamma_0 < \gamma_\text{irreversible} \sim 7\%$~\cite{keim2013,keim2014,keim2015,keim2020}. Therefore, while we begin by decrementing the amplitude $\gamma_0$ in steps of $\Delta \gamma = 1\%$, below $\gamma_0 = 10\%$ we take smaller steps $\Delta \gamma = 0.25\%$, based on the presumption that the material can form memories and so the discreteness of the steps affects the final outcome.

\begin{figure} 
    \centering
    \includegraphics[width=3.4in]{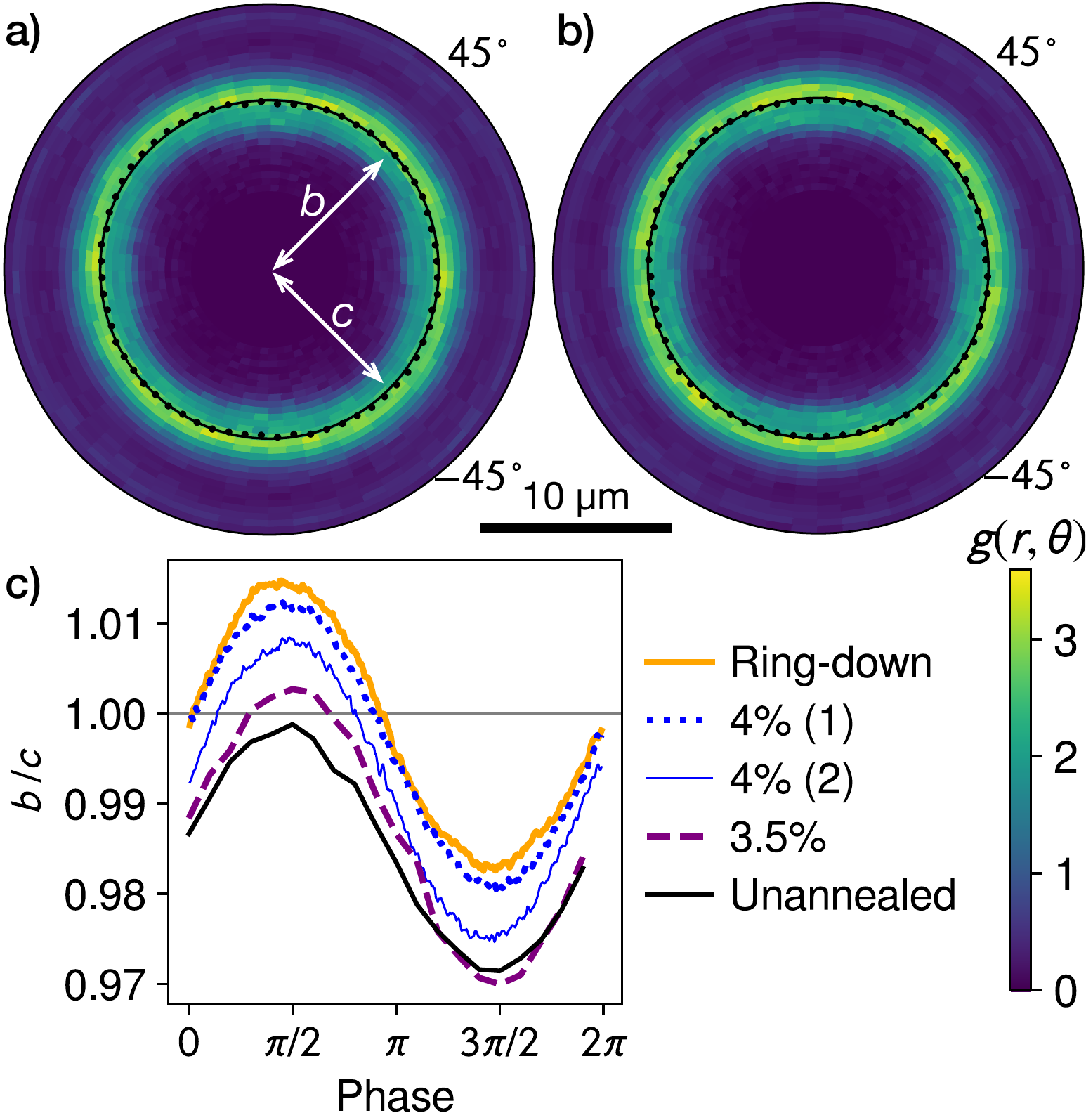}
    \caption{Annealing and microstructure.
    \textbf{(a)} Pair-correlation function $g(r, \theta)$ at end of large-amplitude reset, i.e.\ unannealed. Color bar is at lower-right of figure. Black dots indicate nearest-neighbor peak at each $\theta$. Black circle is for comparison, showing that ring is slightly elongated in the $-45^\circ$ direction. Ellipse fit to the peaks (not drawn for clarity) has semi-axes $b$ and $c$, constrained to the $\pm 45^\circ$ directions.
    \textbf{(b)} Corresponding plot after ring-down annealing, showing no single axis of elongation.
    \textbf{(c)} Ellipse axis ratio $b/c$ during 3.5\%-amplitude cycles after different annealing, showing that ring-down relaxes underlying asymmetry; constant-amplitude annealing at 4\% (two trials) and 3.5\% has smaller, inconsistent effect. Upper three curves are 3-frame rolling averages from experiments with more video frames.}
    \label{fig:asymm}
\end{figure}

\subsection*{Effect of annealing on microscopic structure} 

Both constant-amplitude and ring-down annealing systematically change the material's microstructure. Figure~\ref{fig:asymm} shows pair-correlation functions $g(r, \theta)$ that characterize the average positions of each particle's nearest neighbors, where we divide $r$ into bins of width 0.5~px, and $\theta$ into bins of width $360^\circ / 64$. From continuum mechanics we expect that a positive shear strain $\gamma \ll 1$ would extend the microstructure along the $\theta = 45^\circ$ principal axis and compress it along the $-45^\circ$ axis; an initially circular $g(r, \theta)$ would become an ellipse. To measure this anisotropy, along each azimuthal direction $\theta$ we take the one-dimensional $g(r)$ and find the center of mass of the first peak with sub-pixel resolution. We then fit these peak positions with an ellipse that has semi-axes $b$ along the $45^\circ$ axis and $c$ along the $-45^\circ$ axis.

Intuitively, the condition $b=c$ should correspond to mechanical equilibrium, with no imposed shear stress in either direction. However Galloway et al.~\cite{galloway2020,galloway2021} found that an elliptical signature could be detected even at equilibrium, and that it survived further shear deformations below the yielding transition. They identified it as the memory of an earlier large plastic deformation---consistent with scattering results from bulk metallic glasses that were previously deformed~\cite{ott2008,sun2016a,sun2016}. Teich et al.~\cite{teich2021} showed that the remnant anisotropy was most evident in regions that resisted rearrangements, giving this memory its stubborn persistence. These results prompt the question of how one would ever be truly free of this anistropy once it has formed: large-amplitude shear erases the memory, but might necessarily leave behind another one when it stops.

Our experiments observe this persistent memory of a direction, but also show that annealing can remove it. Figure~\ref{fig:asymm}a shows that immediately after large-amplitude shear, even though the material is ostensibly undeformed ($\gamma = 0$), the quenched microstructure is asymmetric with $b / c < 1$. On the other hand, the ring-down-annealed sample in Fig.~\ref{fig:asymm}b does not show this anisotropy---the peaks are farther along both $45^\circ$ and $-45^\circ$ equally, which is likely an artifact of the camera's pixel grid. To place these observations in context, Fig.~\ref{fig:asymm}c plots the asymmetry $b/c$ over a cycle of shear with amplitude 3.5\%, for an unannealed sample immediately after reset, a sample annealed with constant 3.5\% amplitude for 100 cycles, two samples annealed with constant 4\% amplitude for 128 cycles, and a sample prepared with ring-down annealing. The 4\%-annealed and ring-down samples were observed during the readout protocol.
The asymmetry oscillates with imposed shear, as expected, but annealing can erase the underlying asymmetry of the quenched microstructure, making the oscillation nearly symmetric about $b/c = 1$. 
Moreover, ring-down annealing does this consistently, and not just in the representative curve plotted: across 14 trials, the mean of $b / c$ at zero strain---either just after annealing or during readout---is $0.9993$ with standard deviation $0.0024$.
By contrast, the effectiveness of constant-amplitude annealing seems to depend on the amplitude, and the two trials performed with 4\% amplitude in Fig.~\ref{fig:asymm}c seem much less consistent than the relatively narrow distribution from ring-down annealing.

It is likely that one could improve the constant-amplitude results by choosing a larger amplitude, specifically the critical amplitude that simulations suggest leads to statistically identical structures from disparate initial conditions~\cite{liu2020,bhaumik2019,sastry2021}. However, those same simulations show that just above that amplitude, the response begins to take on the character of irreversible plastic flow, which is how the memory of a direction was formed in the first place. Ring-down annealing avoids this fine tuning.

\begin{figure} 
    \centering
    \includegraphics[width=3.4in]{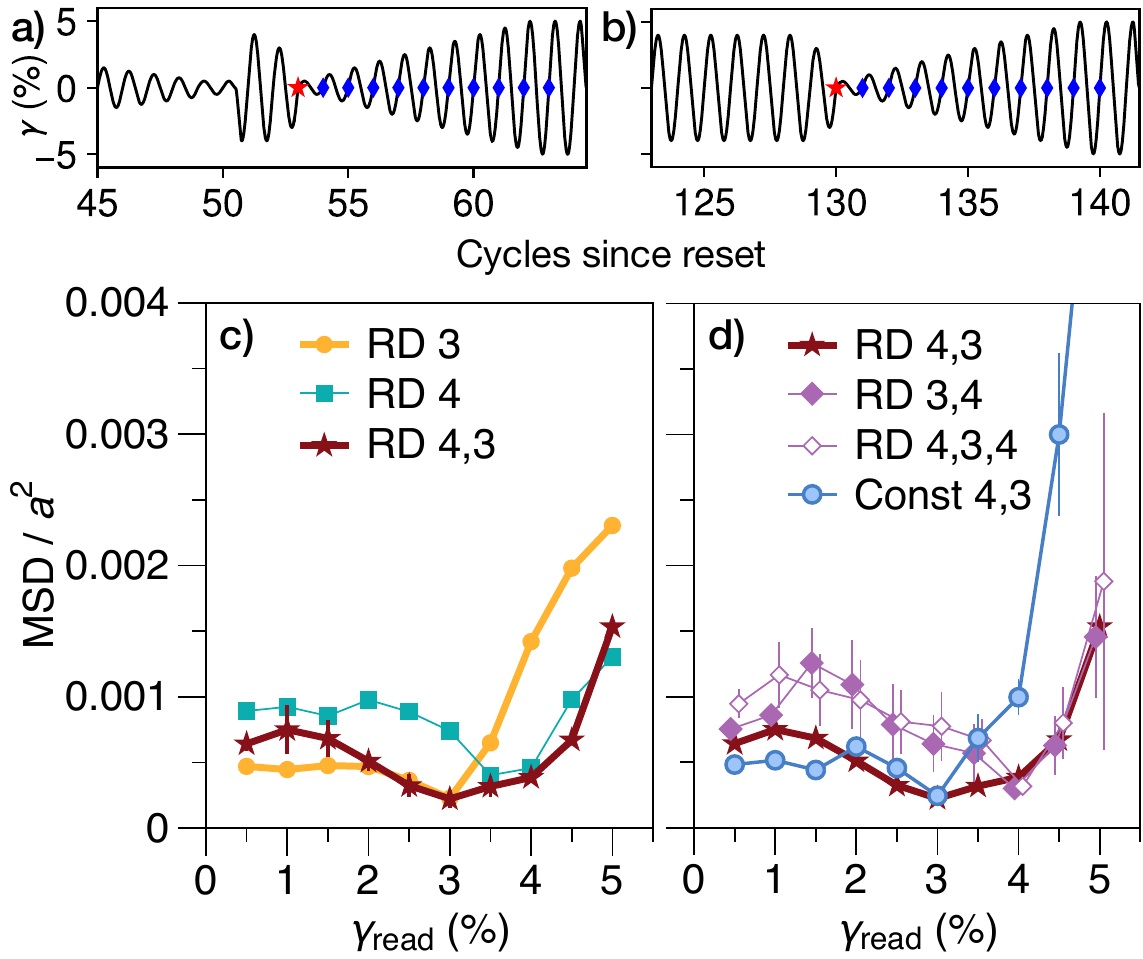}
    \caption{Annealing and nested memories.
    \textbf{(a)} After ring-down, cycles write memories of 4\%, 3\% amplitude before readout (marked as in Fig.~\ref{fig:simple}(b,~c). 
    \textbf{(b)} Alternately, after constant-amplitude annealing writes the 4\% memory, a single cycle encodes 3\% before readout.
    \textbf{(c)} Readout with nested memories from ring-down protocol (a), labeled ``RD 4, 3.'' Single-memory curves ``RD 3'' and ``RD 4'' reproduced from Fig.~\ref{fig:simple}(d). Two-memory curve is distinct: dip indicates memory at 3\%, and change in slope indicates memory at 4\%. 
    \textbf{(d)} Results of alternate two-memory protocols. ``RD 4, 3'' is from (c). Smaller memory is erased by reversing order of write cycles (``RD 3, 4'', offset left for clarity) or inserting a final 4\%-amplitude write cycle (``RD 4, 3, 4'', offset right). ``Const 4, 3'' curve shows that beginning with constant-amplitude annealing (b) also allows two memories, but particle positions are dramatically more disrupted above $\gamma_\text{read} = 3\%$ and especially above $4\%$. Error bars for ``RD 4, 3'' and ``RD 4, 3, 4'' show range of 2 trials; all others show standard deviation of mean for 3 trials.}
    \label{fig:multiple}
\end{figure}

\subsection*{Memories of amplitude} 

In Fig.~\ref{fig:simple}(c, e) we showed that once a system is prepared by ring-down annealing, a single cycle of shear with amplitude $\gamma_1$ writes a memory of $\gamma_1$. (The cycle is the sequence $\gamma = 0 \to \gamma_1 \to -\gamma_1 \to 0$; the preceding excursion to $-\gamma_1$ is added to smoothly start up shear and ensure a complete cycle that begins and ends at $\gamma = 0$.) The form of these readout curves is consistent with the central principle of return-point memory~\cite{barker1983,sethna1993,keim2019}: that as long as the strain is bounded by two turning points (in this case $\gamma_1$ and $-\gamma_1$), returning the strain to either of those values will restore the system to its state when that strain was last visited. In this case, we expect that once the driving has visited $\gamma_1$ during the readout cycle with $\gamma_\text{read} = \gamma_1$, driving the system back to $\gamma = 0$ will retrace the same particle motions as during the writing cycle---making the particle positions match those at the end of writing, and minimizing the MSD~\cite{keim2019,paulsen2019,keim2020}. 

\begin{figure} 
    \centering
    \includegraphics[width=4.5in]{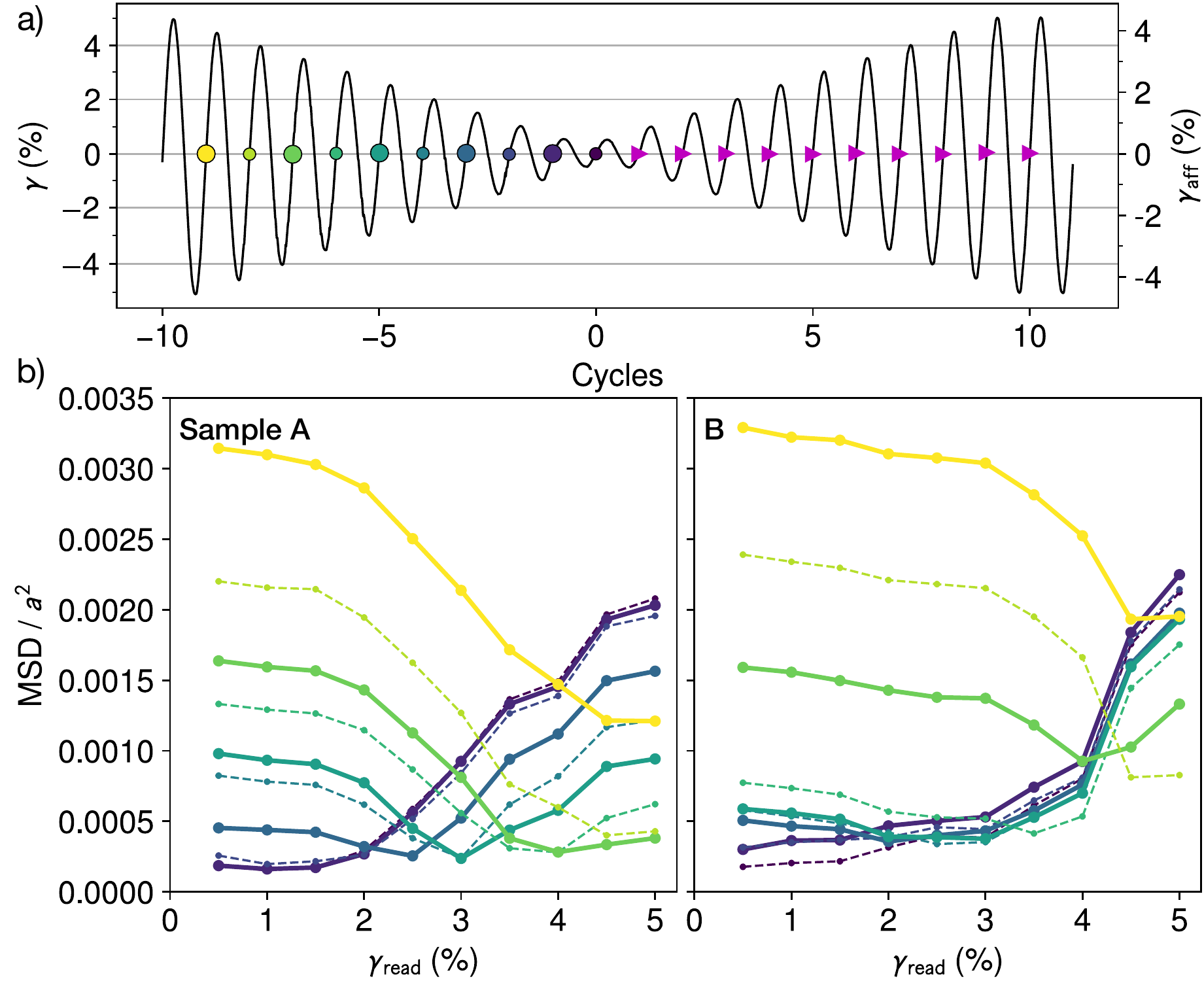}
    \caption{Ring-down as forming nested memories. 
    \textbf{(a)} Ring-down annealing (not shown) is followed by 10 more ring-down ``write'' cycles with decreasing amplitude (ends marked with colored circles), and matching ring-up readout cycles (ends marked with magenta triangles). 
    Left axis shows nominal strain $\gamma$; right axis shows $\gamma_\text{aff}$ measured from particle positions, for comparison.
    \textbf{(b)} Ring-up restores states from ring-down in reverse order. Each colored large (small) circle in (a) corresponds to a solid (dashed) curve which compares that state with the states during ring-up. 
    Darkest dashed curve in each plot refers to end of ring-down; the average of this curve for both samples is in Fig.~\ref{fig:simple}e. Sample A discriminates better among closely-spaced memories.}
    \label{fig:rdru}
\end{figure}

This kind of instantaneous writing is seen in other systems with return-point memory~\cite{barker1983,perkovic1997,paulsen2019,mungan2019}, in contrast to the gradual memory formation in systems with multiple transient memories, including non-Brownian suspensions~\cite{keim2011b,paulsen2014a,paulsen2019}. Prior work on memory readout in disordered solids has written the memories with many repetitions~\cite{fiocco2014,adhikari2018a,mukherji2019,keim2020}, conflating the processes of annealing and writing. Our new experiments show that they may be separated. We can also omit the writing step altogether and perform readout immediately after a ring-down, resulting in the monotonically increasing ``No write'' curve in Fig.~\ref{fig:simple}e. We will more precisely describe this curve and its memory content in the discussion of Fig.~\ref{fig:rdru}.

In Fig.~\ref{fig:multiple}(a, c) we show that after ring-down annealing, return-point memory applies recursively and lets us nest one memory within another. We apply a cycle of 4\% amplitude and then a cycle of 3\%. Because the turning points at $\pm 3\%$ are nested within the previous pair $\pm 4\%$, this second cycle does not reverse the effects of the first. The resulting ``RD 4, 3'' curve in Fig.~\ref{fig:multiple}c shows that a readout cycle with $\gamma_\text{read} = 3\%$ amplitude restores the state from the end of writing, as would also be expected with a single memory. However, the subsequent readout above 3\% fails to change the system as much as in the single-memory case, revealing that at these amplitudes, the system was prepared differently. We expect that the actual memory at 4\% is indicated by an increase in slope of the readout curve~\cite{perkovic1997,keim2019,keim2020}, which is difficult to locate with precision in these data. We also verify the hierarchical nature of this memory, consistent with a return-point description~\cite{barker1983,sethna1993,paulsen2019,keim2020}: memories of multiple values must be nested inside each other by writing in descending order and reading in ascending order (the ``RD 4, 3'' curve in Fig.~\ref{fig:multiple}d). Reversing the order of writing (``RD 3, 4'') results in a single memory, as does overwriting a dual memory before readout (``RD 4, 3, 4'').

Figure~\ref{fig:multiple}d also underscores that while both ring-down annealing (``RD 4, 3'') and constant-amplitude preparation (``Const 4, 3'') allow encoding and readout of the same information, they are not equivalent. Figure~\ref{fig:multiple}b shows how we prepare the material with a constant amplitude of 4\%, then encode a second memory with a single cycle of 3\% amplitude. The greatest difference between these readout curves is for $\gamma_\text{read} > 3\%$. Note that unlike in Fig.~\ref{fig:simple}d, the constant-amplitude data in Fig.~\ref{fig:multiple}d is from the same set of experiments as the ring-down trials, enabling a quantitative comparison.

We note that for the constant-amplitude-annealed samples, readout with amplitude $\ge 4\%$ represents the first time that deformations of this magnitude have been performed since the material was quenched from a yielded, flowing regime (the large-amplitude ``reset'' protocol). Exceeding the envelope of previous deformations should trigger many additional, latent rearrangements. Work by Mungan et al.\ to map out the graph of reachable states under cyclic driving suggests that in general, some of these new rearrangements cannot be reversed by any subsequent deformation; the material is changed permanently~\cite{mungan2019}. This aspect of the material's response is reminiscent of the Mullins or Kaiser effects in other systems, where plastic damage constitutes a simple memory of the largest deformation applied~\cite{keim2019}. By contrast, ring-down annealing subjects the quenched material to many cycles with amplitudes $\ge 4\%$, so that readout with $\gamma_\text{read} > 4\%$ will trigger few (if any) irreversible rearrangements, consistent with its lower MSD signal.

\subsection*{Nested memories from ring-down}

While we have shown that two memories can coexist in this system, in theory, return-point memory allows arbitrarily many values to be stored~\cite{barker1983,sethna1993,paulsen2019,keim2020}.  An obvious way to test this prediction is to check whether $n$ encoded memories are present in readout data, as we did for $n=2$ in Fig.~\ref{fig:multiple}. Such a readout-based approach is important to a complete description of memory capacity, but it presents the additional challenge of establishing a method to extract as many features as possible from experimental readout data. However, a key prerequisite for memory capacity is that the material itself can distinguish among input values, i.e., that applying $n$ different strain amplitudes in descending order results in $n$ distinct states, that can each be recovered later according to return-point memory. This is achieved with a ring-down protocol.

Figure~\ref{fig:rdru}a shows our protocol to test how ring-down forms nested memories. After ring-down annealing, we apply another ring-down series of 10 cycles with amplitudes from 5\% to 0.5\%, in steps of 0.5\%. At the end of each cycle the system is ostensibly in a new state that encodes one more memory. As we apply the corresponding ring-up series for readout, each of those states should be recalled in reverse order. Fig.~\ref{fig:rdru}b tests this for each of two samples. For example, the 4\% curve in each panel (solid green line) compares each state during ring-up to the state after the $4\%$ cycle in ring-down (marked in Fig.~\ref{fig:rdru}a with a circle of the same color). If the material had an unlimited ability to resolve different memories, we would expect each curve to have a global minimum at the corresponding $\gamma_\text{read}$, where the curve would drop sharply to nearly zero MSD. Instead, we see that Sample A in Fig.~\ref{fig:rdru}b distinguishes poorly among strains $\lesssim 2\%$, suggesting a practical lower bound on distinct memory values. We also see that the 5\% readout curve for Sample A (solid dark blue) has a large minimum MSD, meaning that that this memory state---the first to be recorded and last to be retrieved---was not recalled faithfully, and hinting at a practical upper bound on memory values. Between these bounds, the system further struggles to discriminate between 3.5\% and 4\%, indicating that they are too close to be distinct.

The right panel of Figure~\ref{fig:rdru}b represents the same test performed on Sample B, which is the same material, but with a differently randomized structure due to the reset protocols and several other intervening tests. We again see signs of insufficient memory capacity, but now the deficiency is more severe, extending up to $\gamma_\text{read} \approx 3\%$. These two samples show that the ability of the $\sim$1400 rearranging particles to store and distinguish among strain amplitudes is surprisingly limited, and surprisingly variable. 

\begin{figure} 
    \centering
    \includegraphics[width=\textwidth]{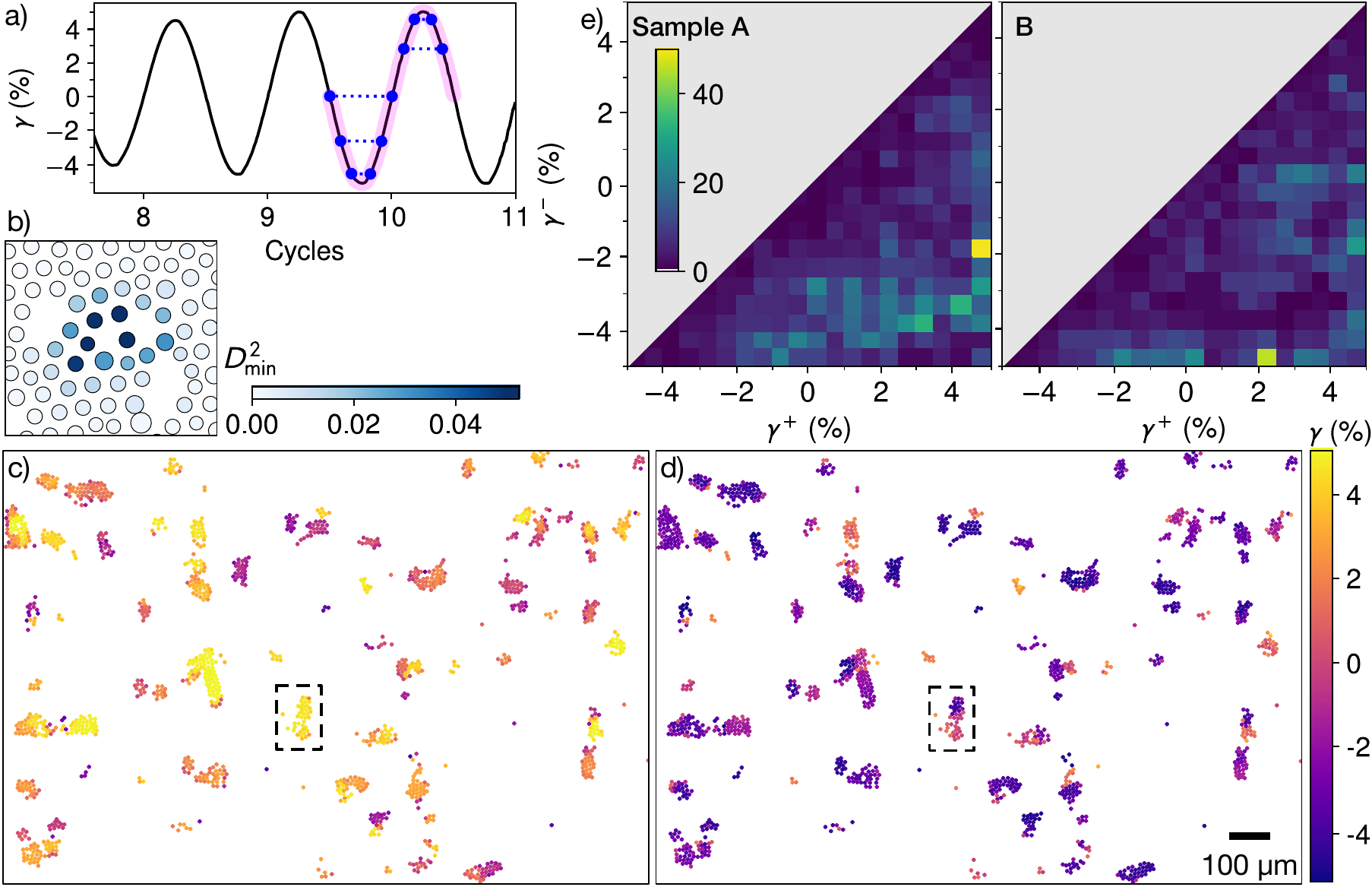}
    \caption{Spatiotemporal structure of rearrangements within a single cycle.
    \textbf{(a)} Sampling method. A cycle with amplitude $5\%$ (highlighted pink), in which particle trajectories are closed, is taken from the end of a readout protocol. Five representative pairs are marked with blue dots and dotted lines, matching each frame during forward shear ($\dot \gamma > 0$) with the frame at same $\gamma$ during reverse shear.
    \textbf{(b)} Squared non-affine displacement $\dtwomin$ of particles at the center of the soft spot in Fig.~1a. Voids represent particles that could not be reliably tracked (see text). 15 particles shown had values exceeding the threshold 0.015. Largest value is 0.14; color scale is clipped to show small values clearly.
    \textbf{(c, d)} Matching panels showing all rearranging particles in the field of view, colored by the strains $\gamma^+$ ($\gamma^-$) at which each rearranges during forward (reverse) horizontal shear. Fixed wall is at the top; moving needle is at the bottom. Dashed outline highlights example suggesting two strongly coupled soft spots.
    \textbf{(e)} Histograms of $\gamma^+$, $\gamma^-$ in two identically prepared material samples, A and B. Sample A corresponds to (c, d). Color indicates number of particles in each bin.}
    \label{fig:stzmap}
\end{figure}

\subsection*{Mapping rearrangement kinematics} To explain the limitations revealed in Fig.~\ref{fig:rdru} we return to localized plastic rearrangements like the one in Fig.~\ref{fig:simple}a. The loci of these rearrangements are often termed ``soft spots'': mesoscopic regions that are predisposed to rearrange under shear, playing the role of shear-transformation zones~\cite{falk2011,manning2011,cubuk2017}.  Experiments by Keim at al.~\cite{keim2020} and simulations by Regev, Mungan et al.~\cite{mungan2019,regev2021} showed how return-point memory arises from the hysteresis of many soft spots. Here, we develop a way to efficiently map the kinematic structure of these soft spots, before considering how it connects with return-point memory. Following earlier analyses, we model each soft spot as a bistable ``hysteron'' with states $+1$ and $-1$~\cite{falk2011,manning2011,keim2014,perchikov2014,mungan2019,keim2020}. Under forward shear ($\dot \gamma > 0$), the $i$th soft spot transitions to the $+1$ state when $\gamma > \gamma^+_i$; under reverse shear it transitions to the $-1$ state when $\gamma < \gamma^-_i$. The thresholds $\gamma^\pm_i$ vary among soft spots to represent the disorder of the system. We also require $\gamma^+_i > \gamma^-_i$ to represent that each soft spot is dissipative---the rearrangement uses stored elastic energy. This also makes each soft spot hysteretic: when $\gamma^-_i \le \gamma \le \gamma^+_i$, the state of the $i$th soft spot depends on its history. 

We apply the hysteron model to experiments by measuring the individual $\gamma^\pm_i$ of the many soft spots that rearrange and reverse during one cycle of shear with amplitude $\gamma_0$i.
The simplest method is to start at a chosen time $t_0$ and measure particle displacements from that instant. However in practice this tends to be more sensitive to rearrangements at strains far from $\gamma(t_0)$, as was evident in a previous measurement~\cite{keim2014}. To construct an unbiased method, we note again that when $\gamma_i^- < \gamma < \gamma_i^+$, the $i$th soft spot can be found in either state, depending on the direction of shear. Checking for the direction-dependence of each soft spot, at many values of $\gamma$ during a full cycle in which particle trajectories are closed, will thus reveal the values of $\gamma_i^-$ and $\gamma_i^+$.

In our experiments we take advantage of the two-cycle segment with $\gamma_0 = 5\%$ at the end of each readout (e.g.\ Fig.~\ref{fig:rdru}a). We avoid using the very end of this segment---the last frame of the movie---since the strain may not return fully to $\gamma = 0$ before video recording ends. We instead select a full cycle from the middle of this segment, highlighted in Fig.~\ref{fig:stzmap}a. To the extent that the system follows return-point memory, this cycle will leave the material unchanged, since the immediately preceding visit to $\gamma = 5\%$ restored the system to the corresponding state during ring-down, and erased any smaller memories. Each video frame during forward shear ($\dot \gamma > 0$) is matched with its counterpart at the same $\gamma$ during reverse shear. We compute each particle's squared non-affine displacement $\dtwomin$ between this pair of frames (see Materials and Methods), and take particles with $\dtwomin \ge 0.015$ as being in different states---a threshold we established previously as reliably isolating rearranging regions~\cite{keim2014,keim2015}, and that does not have a strong qualitative effect on our results (see Supplemental Information). We reject any particle that did not have a closed trajectory during the cycle (no particles in the case of Fig.~\ref{fig:stzmap}[c, d]), or whose $\dtwomin$ is elevated for fewer than 2 frames, or for less than half of the frames between its $\gamma^+$ and $\gamma^-$. 

Figures~\ref{fig:stzmap}(c, d) show the rearranging particles in Sample A, colored by the $\gamma^+$ and $\gamma^-$ we obtain. Consistent with return-point memory, all soft spots returned to their original states at the end of the cycle (Sample B had two small exceptions; see Supplemental Information). We see that neighboring particles tend to have similar values, and there is often very little variation between the core of a small soft spot and its periphery, confirming that the switching of a single soft spot may be treated as a discrete event, and that our experiments are approximately quasistatic. We also see that some extended regions that might appear as a single soft spot in a map of e.g.\ $\dtwomin$, actually have two or more distinct values of $\gamma^+$ or $\gamma^-$, indicating that they should be treated as separate hysterons in our model. Such locations are opportunities to study small groups of strongly coupled soft spots~\cite{mungan2019,keim2021,lindeman2021a}. Indeed, in many cases these interactions are apparent: for example, the region outlined in Fig.~\ref{fig:stzmap}d rearranges in two stages during reverse shear, with varied $\gamma^-$, but all at once during forward shear, with nearly uniform $\gamma^+$---an avalanche behavior identified in the simulations of Mungan et al.~\cite{mungan2019} that is explained by a cooperative (i.e.\ ferromagnetic) coupling.

In Fig.~\ref{fig:stzmap}e we plot the particles' $\gamma^+$ and $\gamma^-$ as histograms on the $\gamma^+$-$\gamma^-$ plane, for Samples A and B. The bin boundaries of the histograms are the strain amplitudes in our memory capacity tests. We see large differences in extended portions of these histograms, just as in Fig.~\ref{fig:rdru} we established that memory capacity can vary between samples. We can now consider whether these two kinds of observations are related.

\begin{figure} 
    \centering
    \includegraphics[width=\textwidth]{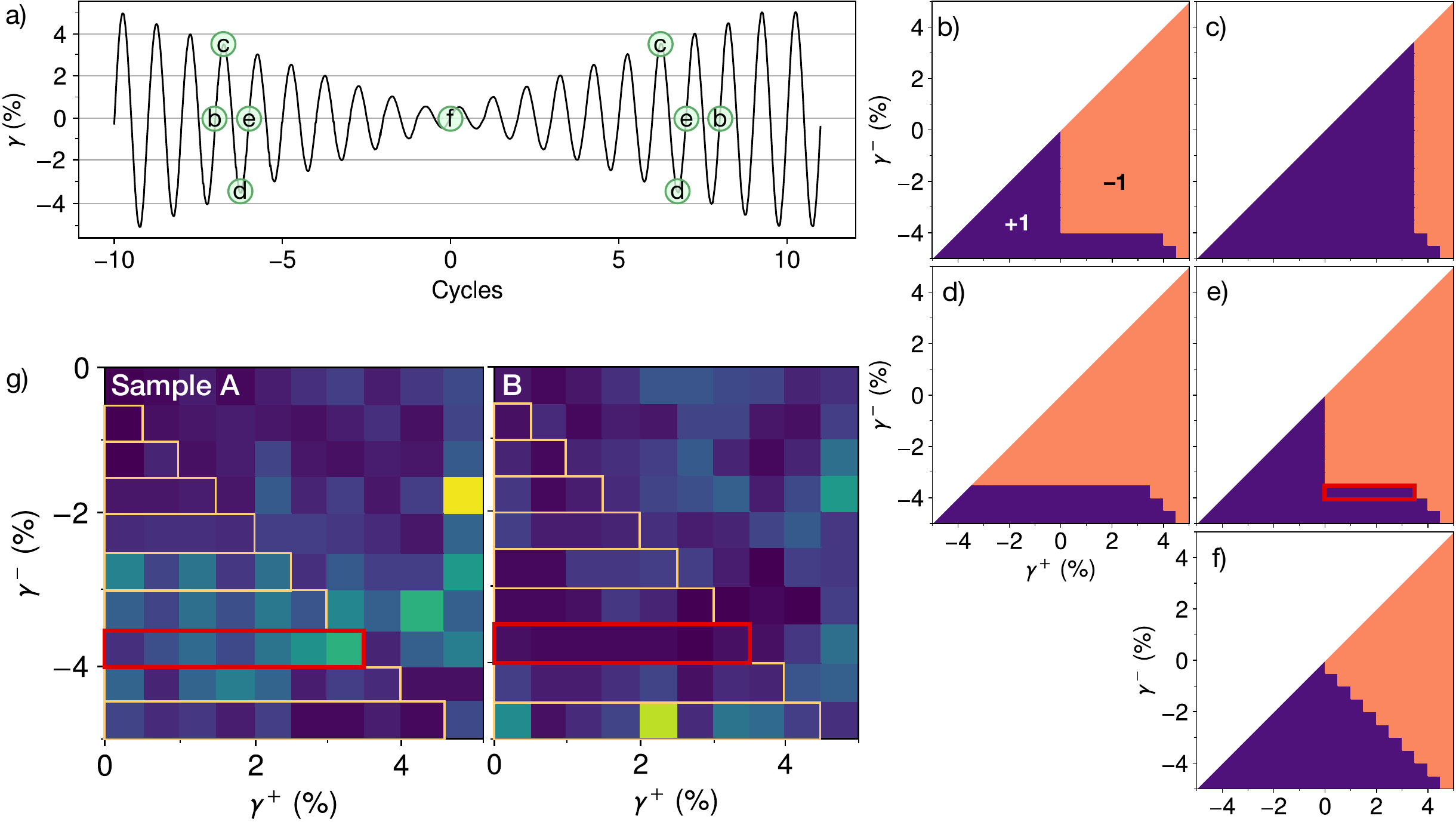}
    \caption{Preisach model of response to amplitude variation.
    \textbf{(a)} Nested memory protocol as in Fig.~\ref{fig:rdru}. Labels b--e mark pairs of identical states, as predicted by return-point memory.
    \textbf{(b--f)} Diagrams corresponding to labels in (a), showing states of many idealized soft spots, distributed uniformly on the $\gamma^+$-$\gamma^-$ plane. (b) is at $\gamma = 0$ immediately following a cycle with amplitude 4\%. To evolve this model to $\gamma = 3.5\%$ at (c) via forward shear, we ensure that every soft spot with $\gamma^+ \le 3.5\%$ has rearranged into its ``$+1$'' state, shaded dark purple. Reverse shear to (d) switches every soft spot with $\gamma^- \ge 3.5\%$ to the ``$-1$'' state. Returning to $\gamma = 0$ at (e) reveals the strip of soft spots with the ``$+1$'' state (outlined in red) that distinguish between 4\% and 3.5\% strain amplitude. Repeating the process with decreasing amplitude leads to (f); increasing amplitude revisits states (b--e).
    \textbf{(g)} Portions of histograms from Fig.~\ref{fig:stzmap}e. Outlined strips show soft spots that let return-point memory discriminate between similar strain amplitudes, as tested in Fig.~\ref{fig:rdru}; red outline matches that in (e). Sample B result suggests relatively poor performance below $\sim$4\%, consistent with Fig.~\ref{fig:rdru}b.}
    \label{fig:capacity}
\end{figure}

\subsection*{Predicting response to variable-amplitude shear}

To predict the response of the soft spots in Fig.~\ref{fig:stzmap}e to changes in strain amplitude, we adopt the simplifying approximation that soft spots do not influence each other, meaning that the $\gamma_i^\pm$ are constant. This makes our model of many soft spots equivalent to the model of magnetic hysteresis developed by Preisach~\cite{preisach1935} that is proven to have return-point memory~\cite{barker1983,sethna1993}; in fact, our annealing method was inspired by the degaussing method for minimizing the remnant magnetization of a ferromagnet. In Figs.~\ref{fig:capacity}(a--f) we consider how ring-down and ring-up operate on this model. Instead of the experimental histograms in Fig.~\ref{fig:stzmap}e, in Figs.~\ref{fig:capacity}(b--f) we assume that soft spots are continuously distributed on the $\gamma^+$-$\gamma^-$ plane, and we represent their \emph{states}. Labels on the strain protocol in Fig.~\ref{fig:capacity}a refer to these diagrams. Figures~\ref{fig:capacity}(b--e) are snapshots from one cycle during ring-down. As we shear forward from $\gamma = 0$ to $\gamma = 3.5\%$, we evolve the system from state (b) to state (c) by sweeping rightward on the horizontal axis from 0 to $3.5\%$, switching soft spots to the $+1$ state as we go. To complete the cycle, we shear from $3.5\%$ (c) to $-3.5\%$ (d) by changing states to $-1$ as we move downward on the vertical axis, and then we perform forward shear back to $\gamma = 0$ (e). The stair-step pattern in Figure~\ref{fig:capacity}f is the result of the complete sequence of ring-down cycles.

Comparing states (b) and (e), we see that soft spots within one rectangular region on the $\gamma^+$-$\gamma^-$ plane (outlined in red) are responsible for storing the new memory of $3.5\%$ amplitude and distinguishing it from the retained memory of 4\%. No matter how many soft spots are active within a material, if is has none with $0 < \gamma^+_i < 3.5\%$ and $-4\% < \gamma^-_i < 3.5\%$ it cannot make this distinction---the 3.5\% cycle will leave the material unchanged.
This lets us interpret the histograms from Fig.~\ref{fig:stzmap}e, portions of which are reproduced in Fig.~\ref{fig:capacity}g. We outline and zoom in on the 9 regions that encoded the memories from 0.5\% to 4.5\% in our model. 
Some regions are relatively deficient in particles, allowing us to predict features of each sample's response. While both samples appear to have poor capacity at $\gamma \lesssim 2\%$, Sample A should also perform poorly around $\gamma \simeq 4.5\%$, while Sample B should perform poorly around $\gamma \simeq 3\%$. These features are consistent with the deficiencies we observed independently in the variable-amplitude tests of Fig.~\ref{fig:rdru}. Given the magnitude of this variation, limited soft spot populations also partly explain the differences among memory readout curves when repeating trials under identical conditions, as shown by the error bars in Figs.~\ref{fig:simple}d and \ref{fig:multiple}(c, d).

The success of our Preisach model analysis suggests that even though our data show that nearby soft spots interact, and there may also be small strain rate-dependent effects that nonuniformly change the timing of rearrangements (see Supplemental Information), the net result of these effects is perturbative in practice, and does not affect the broad features of a sample's response to amplitude variation. We also note that the visible interactions in Figs.~\ref{fig:stzmap}(c, d) are cooperative---causing one rearrangement to trigger another at the same $\gamma^+$ or $\gamma^-$---and so are still formally consistent with return-point memory~\cite{sethna1993}, if not with the Preisach model.

\section*{Discussion}

We have shown through experiments that while conventional mechanical annealing with a constant strain amplitude leaves an imprint of that amplitude, annealing with a decreasing amplitude appears to be similarly effective, while erasing the structural anisotropy from earlier large deformations and leaving no strong signature of a strain. 

Our results also confirm that mechanical annealing of disordered solids does not just reach a steady state, but it imparts a remarkable kind of stable plasticity---so that to a good approximation, plastic changes in the material can be predictably undone by further deformations. This can be described as return-point memory, and our experiments test it in depth---first by reading out one or two memories (Figs.~\ref{fig:simple} and \ref{fig:multiple}), and then by nesting 10 memories together (Fig.~\ref{fig:rdru}). These experiments show the excellent fidelity of return-point behavior within a $\sim$8\% window of strain ($-4\% < \gamma_\text{aff} < 4\%$). The ability to undo plastic deformations seems at odds with the basic nature of disordered solids: the disordered structure of these materials is not quenched, and the soft spots that control plasticity can be created and destroyed~\cite{falk2011,manning2011}; there is no guarantee that the soft spots that couple to shear (Fig.~\ref{fig:stzmap}) will remain from one cycle to the next as we vary the amplitude~\cite{mungan2019,keim2020}. As a counter-example, shearing the un-annealed material twice with the \emph{same} amplitude generally yields two different states (e.g.\ \cite{regev2013a,keim2013}). 

Recent simulations~\cite{yeh2020,bhaumik2019,sastry2021,liu2020} that measure a sample's structural energy---the total potential energy of interparticle interactions---hint at one underlying reason that ring-down annealing can erase the past and achieve this kind of stability. For materials such as ours that yield by flowing homogeneously (rather than with a shear band), ring-down passes through a critical point at the material's yielding amplitude. \emph{Constant}-amplitude shear near this critical point erases a sample's thermal history by approaching a critical energy per particle---the lowest achievable energy for such a material in these athermal studies, which requires exploring a large number of possible configurations~\cite{szulc2020,regev2021}. In practice, it is unclear how small $\Delta \gamma$ should be near the critical point to achieve this outcome. Simulations have shown that annealing a material to a steady state at a strain amplitude just below this point may require hundreds or even thousands of cycles~\cite{regev2013a,lavrentovich2017,bhaumik2019}, calling for vastly slower annealing than we have demonstrated here. However, these and other studies also show that most of the relaxation takes place early in the process, so that a much smaller number of cycles may suffice. Greater efficiency might also be possible with more sophisticated, non-monotonic ways of varying amplitude near the critical point~\cite{priezjev2021}.
Once the system has been allowed to evolve near the critical point, subsequently lowering the strain amplitude may achieve still-lower energies, as in experiments that ramp down driving of a granular packing to approach a structure with the lowest accessible energy~\cite{nowak1997,nicolas2000}. Simulations with a ring-down protocol might therefore reveal more of this energetic picture.

Our results grew out of the study of memory formation, and they are an invitation to explore connections between memory and the ways that annealing affects stability and configurational entropy. We posit that a discernible memory---here, of an amplitude or direction---is one sign of incomplete mechanical annealing. On its face this contradicts our return-point memory tests, which suggested that the ring-down protocol actually writes many finely-spaced memories in the material. To resolve this tension, we consider the spacing of strain amplitude steps during ring-down annealing, $\Delta \gamma$.
In our analysis we modeled soft spots as non-interacting (a Preisach model; Fig.~\ref{fig:capacity}), which let us understand a particular sample's limited ability to distinguish among closely-spaced memories (Fig.~\ref{fig:rdru}) in terms of its finite population of soft spots (Fig.~\ref{fig:stzmap}).  
In an idealized sample with an unlimited number of soft spots, repeatedly halving $\Delta \gamma$ would subdivide each stair step in Fig.~\ref{fig:capacity}f, resulting in an ever-finer structure of nested memories. However, in a finite material there may not be enough soft spots to encode these finer steps. Eventually, decreasing $\Delta \gamma$ below some value $\Delta \gamma_\text{min}$ will make no further difference in the outcome of annealing. Upon continuing to $\Delta \gamma \ll \Delta \gamma_\text{min}$, readout will be dominated by the granularity of the soft spots, and the written memories will be indistinct. In this limit, the states of the soft spots encode only the equilibrium strain $\gamma = 0$ around which shearing was symmetric, and the fact that a ring-down protocol was employed---no matter how sensitive the readout technique. 
Because $\Delta \gamma_\text{min}$ restricts the number of retrievable memories, it is connected with the memory capacity of the system. We can therefore use the simulation results of Regev et al.~\cite{regev2021} (or the area of each square stair-step in Fig.~\ref{fig:capacity}f) to suggest that $\Delta \gamma_\text{min}$ scales as $1/\sqrt{N}$, where $N$ is the number of particles. Paradoxically, the most gradual form of ring-down annealing thus achieves a ``memory-free'' state by saturating the memory capacity of the system. This prepares the system to record any subsequent deformation, by reverting to an earlier state without the smallest nested memories---a change that can be read out (Fig.~\ref{fig:simple}e).

By carefully studying the relationship between annealing and memory, we have also shown that the complex multiple-memory behavior that was previously observed after many cycles of ``training''~\cite{fiocco2014,adhikari2018a,mukherji2019,keim2020}---carried over from studies of dilute suspensions~\cite{keim2011b,paulsen2014a}---is actually the superposition of two distinct types of memory. This result brings to mind studies that show how various memory behaviors across many disparate systems belong to a much smaller set of distinct types, each with characteristic rules for encoding, reading, and erasing~\cite{keim2011b,keim2019,paulsen2019}. In this case, shearing with a constant amplitude or a repeated pattern of amplitudes anneals the material and forms a memory of the largest deformation applied, revealed by the damage that occurs rapidly when that deformation is exceeded (Figs.~\ref{fig:simple}d, \ref{fig:multiple}d, and Refs.~\cite{benson2020,arceri2021a})---resembling the Mullins or Kaiser effects seen in other materials~\cite{keim2019,paulsen2019}. Once annealed, the sample also allows the memory of multiple strains to be written in single cycles, consistent with return-point memory that was first studied in ferromagnets~\cite{barker1983}. While the way that soft spots give rise to return-point memory is now established~\cite{keim2020,mungan2019,regev2021}, there is likely still another, higher-order kind of memory arising from their frustrated (i.e.\ antiferromagnetic) interactions~\cite{lindeman2021a}. Lastly, these types of memory rooted in a fixed population of soft spots seem inadequate to capture this material's additional memory of the direction of steady shear~\cite{teich2021,galloway2021,patinet2020}.
Placed among the great variety of systems with memory, a disordered solid is a chimera. 

Our experiments point to an effective, purely mechanical way to prepare disordered solids into a known state with reversible plasticity, minimal anisotropy and memory content, a maximal ability to form new memories, and a structural energy approaching the lowest mechanically-accessible value. An area where this method could be especially valuable is rheometry of soft glassy solids, including concentrated emulsions~\cite{macosko1994,kim2017}: besides the need to exclude memories of past deformations, one may also wish to suppress the transient response that follows each change in strain amplitude. To the extent that annealing can ensure a relaxed structure with return-point memory behavior, the transients at the start of each test will be as short as possible; the second cycle after a change in amplitude will be exactly like every one that follows~\cite{keim2011b,lindeman2021a}. More broadly, the memory-based perspective behind the ring-down method also leads to ways of characterizing the population of plastic rearrangements in a sample, solely by comparing observations taken at the same strain---suggesting that our $\dtwomin$ calculations could be replaced by image subtraction (see Supplemental Information) in experiments where single-particle tracking is impractical, and that our methods could even be generalized to very different glassy materials such as crumpled sheets~\cite{shohat2021}. We expect that the family of techniques we describe will allow researchers to mechanically probe and manipulate the rich non-equilibrium character of many materials and systems.


\section*{Materials and Methods}
\textbf{Material samples ---} 
The material shown in Fig.~\ref{fig:simple}a consists of polystyrene sulfate latex particles (Invitrogen), with diameters 4.1~$\mu$m (Lot 1876103) and 5.4~$\mu$m (Lot 1818113) in roughly equal numbers (Fig.~\ref{fig:simple}a). These particles are adsorbed at the interface between decane (``99\%+,'' ACROS Organics) and deionized water in a 60~mm-diameter glass dish~\cite{keim2014}. The particle suspension, with 50\% ethanol as a spreading agent, is handled using pipette tips and Eppendorf tubes that are free of surface treatments (Axygen ``Maxymum Recovery''). 
The typical spacing between particle centers is $a=8.8$~$\mu$m, corresponding to the first peak of the radial pair correlation function $g(r)$~\cite{trackpyv042}. In some movies we observe voids and small aggregates that are presumably due to trace contaminants, but in experiments these act as rigid inclusions and have no special role in rearrangements or memory (see Supplemental Information).

Constant-amplitude data in Fig.~\ref{fig:simple}(d) are from experiments described in Ref.~\cite{keim2020}, with a similar material, apparatus, and analysis methods as in this work; they are presented here for qualitative comparison. 

\textbf{Area fraction measurement ---} We follow the procedure from Ref.~\cite{keim2020}. After applying a short-pass filter to remove background variations, and cropping the image to remove boundary effects, we find the highest grayscale threshold that preserves small particles, and the lowest grayscale threshold that does not merge neighboring particles. We fill holes in the resulting binary images and measure the fraction of dark pixels in each, which gives us a range of area fractions for our estimate.

\textbf{Apparatus ---} The interface is pinned at the edges of two aluminum walls that form an open-ended channel 18~mm long and 2.4~mm wide. A 32~mm-long magnetized steel needle of diameter 233~$\mu$m is adsorbed at the interface in the channel. The needle is trapped at the center of the channel---parallel to the walls---by a pair of permanent magnets suspended above its ends, following the design of Tajuelo et al.~\cite{tajuelo2016,qiao2021}. A translation stage (Physik Instrumente L-509 stage and C-884 controller) moves the magnets parallel to the channel, driving the needle and shearing the material between the needle and the walls. Compared with traditional ISRs that use stationary Helmholtz coils as a magnetic trap\cite{shahin1986,brooks1999,reynaert2008}, it is easier to achieve strong field gradients that tightly couple the needle to the trap, in order to better control strain for the present experiments. To minimize misalignment of the needle, the apparatus is oriented in the horizontal plane to match the local background magnetic field in the lab. In-house software and electronics control the motion and synchronize it with video capture.

\textbf{Shear deformation ---} We achieve a nominal strain amplitude $\gamma_0$ by displacing the magnetic trap with amplitude $\gamma_0 R$, where $R$ is the gap between the needle and each wall. Because of the elastic modulus of the material and the finite stiffness of the magnetic trap, the actual motion of the needle is slightly smaller than $\gamma_0 R$. We measure this actual strain by a least-squares fit of an affine transform to all particle motions, and denote it $\gamma_\text{aff}$. 
When needed, the mapping from these discrete values to $\gamma_\text{aff}$ is given by the extrema of strain during the readout protocol at the end of each experiment. In Fig.~\ref{fig:rdru}a we also show $\gamma_\text{aff}$ for comparison.

Shear is nearly quasistatic, in that the typical timescale for a rearrangement to complete ($\sim$3~s) is much shorter than both the period of shearing (20~s) and the inverse of the maximum strain rate during memory experiments (64~s). Shear is also nearly uniform, but at that maximum strain rate we detect slight non-uniformities because the interface is weakly coupled to bulk viscous flows in the oil and water, which have a different velocity profile~\cite{reynaert2008}. In the Supplemental Information we analyze this behavior and show that the effect is small: when the needle is moving the fastest (i.e.\ near $\gamma = 0$ when $\gamma_0 = 5\%$) the non-uniformity advances or retards the local shear strain by at most 0.2\% strain relative to the global value, depending on location.

\textbf{Particle tracking ---} During deformation, we image $\sim$24,000 particles in an area that includes the needle and one wall. We use a long-distance microscope (Infinity K2/SC) and 4-megapixel machine-vision camera (Mikrotron 4CXP) at a magnification of 0.665~$\mu$m/pixel and a  rate of 20 frames/s. High-throughput tracking is performed with the open-source ``trackpy'' software~\cite{trackpyv042,crocker1996} using the channel-flow prediction and adaptive search features. An image-registration algorithm assists tracking by measuring occasional global displacements of particles due to external vibrations of the microscope; subsequent analysis is insensitive to these global motions. To reduce the effect of spurious rearrangements caused by particle-tracking errors, we discard any particle that is not tracked continuously over an entire set of samples, e.g.\ the entire readout process.

\textbf{Particle displacements ---} To find a particle's displacement between two video frames for Fig.~\ref{fig:simple}a and MSD measurements, we subtract the average motion of the surrounding material within radius $R_\text{disp}$. This avoids spurious signals due to small motions of the camera or variation of the needle position, yielding $\Delta \vec r_\text{local}$~\cite{keim2014,philatracksv02}. We do not compute displacements of particles within $R_\text{disp}$ of an edge of the field of view.
To compute MSD we use $R_\text{disp} = 16.5a$; in Fig.~\ref{fig:simple}a only we use $R_\text{disp} = 12.5a$ to let us observe particles near the edge of the field of view. We found previously that choosing $R_\text{disp} = 4.5a$ or $8.5a$ did not change our qualitative results~\cite{keim2020}. 

\textbf{Identifying rearranging particles ---} While particle displacements allow us to observe rearranging soft spots and compute global differences, we also wish to identify the particles directly involved in a rearrangement, excluding its extended displacement field. We use the $\dtwomin$ measure developed by Falk and Langer~\cite{falk1998,falk2011}. Given particle positions at two instants, we consider each particle and its two ``shells'' of nearest neighbors (within radius $\sim 2.5a$), and find the best-fit affine transformation tensor that describes their displacements~\cite{philatracksv02}. $\dtwomin$ is the mean squared residual from this fit, normalized by $a^2$. Figure~\ref{fig:stzmap}b shows $\dtwomin$ corresponding to the rearrangement in Fig.~\ref{fig:simple}a.

In rare cases, particles with irregular shapes or sizes can be tracked, but not located with sufficient consistency---so that a single particle may be briefly and erroneously displaced by $\mathcal{O}(1)$~px relative to its neighbors. The result is a solitary particle with a high value of $\dtwomin$---very unlike the region of elevated $\dtwomin$ that indicates an actual rearrangement of several particles. An effective de-noising method is to remove each particle with $\dtwomin$ greater than 5 times the median of its neighbors' values (i.e.\ particles closer than $1.5a$), and then recompute $\dtwomin$ for all remaining particles. This procedure removes few particles in actual rearrangements---for example, only one such particle is missing from Fig.~\ref{fig:stzmap}b.

\bibliography{references-misc,references-synced}
\bibliographystyle{ScienceAdvances}

\noindent \textbf{Acknowledgements:}
We thank Muhittin Mungan, Joseph Paulsen, Ling-Nan Zou, and Larry Galloway for valuable discussions. Muhittin Mungan suggested the name of the ring-down protocol. We thank Kevin Thompson for help in fabricating the shear channel. \\
\noindent \textbf{Funding:} This work was supported by National Science Foundation Grant 1708870.\\
\noindent \textbf{Author Contributions:} DM and NCK designed the apparatus and wrote its control software; DM built, improved, and validated it. NCK designed the study, performed and analyzed experiments, and wrote the manuscript.\\
\noindent \textbf{Competing Interests:} The authors declare that they have no competing interests.\\
\noindent \textbf{Data and materials availability:} 782 experimental images sufficient to verify this paper's results are deposited at \href{https:/dx.doi.org/10.5281/zenodo.5772691}{https:/dx.doi.org/10.5281/zenodo.5772691}. See the Supplemental Information for details. Analysis code for this paper is available upon request.

\end{document}